# Bumblebees minimize control challenges by combining active and passive modes in unsteady winds

Sridhar Ravi[1,2*], Dmitry Kolomenskiy[1*], Thomas Engels[3,4], Kai Schneider[3], Chun Wang[2], Joern Sesterhenn[4] and Hao Liu[1]

[1]Graduate School of Engineering, Chiba University, Japan

[2]School of Aerospace Mechanical and Manufacturing Enigneering, RMIT University, Australia

[3]M2P2-CNRS & Aix-Marseille Université, 38 rue Joliot-Curie, 13451 Marseille cedex 20 France

[4]ISTA, Technische Universität Berlin, Müller-Breslau-Strasse 12, 10623 Berlin, Germany

*Equal contributions

**Abstract**

The natural wind environment that volant insects encounter is unsteady and highly complex, posing significant flight-control and stability challenges. Unsteady airflows can range from structured chains of discrete vortices shed in the wake of an object to fully developed chaotic turbulence. It is critical to understand the flight-control strategies insects employ to safely navigate in natural environments. We combined experiments on free flying bumblebees with high fidelity numerical simulations and lower-order modeling to identify the salient mechanics that mediate insect flight in unsteady winds. We trained bumblebees to fly upwind towards an artificial flower in a wind tunnel under steady wind and in a von Kármán street (23Hz) formed in the wake of a cylinder. The bees displayed significantly higher movement in the unsteady vortex street compared to steady winds. Correlation analysis revealed that at lower frequencies (<10Hz) in both steady and unsteady winds the bees mediated lateral movement with body roll - typical casting motion. At higher frequencies in unsteady winds there was a negative correlation between body roll and lateral accelerations. Numerical simulations of a bumblebee in similar conditions permitted the separation of the passive and active components of the flight trajectories. Comparison between the free-flying and 'numerical' bees revealed a novel mechanism that enables bees to passively ride out high-frequency perturbations while performing active maneuvers and corrections at lower frequencies. The capacity of maintaining stability by combining passive and active modes at different timescales provides a viable means for volant animals and machines to tackle the control challenges posed by complex airflows.

**Significance**

Laboratory studies of insect flight typically employ still air or steady flow; natural flows are unsteady and turbulent. This raises fundamental questions about how insects control flight in realistic environments. We used a combination of experiments on freely flying bees, high-fidelity numerical simulations and reduced order modeling to disentangle the active and passive components of the flight dynamics of bumblebees. In unsteady winds bumblebees perform active maneuvers at lower frequencies while interacting passively with wind unsteadiness at higher frequencies. Our findings reveal a novel flight-control mechanism implemented by bumblebees to cope with complex winds, one that could be widespread among volant organisms and suitable for artificial flying systems operating in outdoor environments.

**Introduction**

Volant insects are able to maintain stable flight in natural environments, despite frequently encountering unpredictable, complex airflows. While laboratory studies of insect flight typically employ steady flow, natural flows are often unsteady, containing de-stabilizing vortices and gusts over a wide range of spatial and temporal scales. These unsteady flows are generated by the interaction between wind and obstacles (1, 2), which results in chains of discrete vortices in the near wake of trees, branches and flowers, and fully developed, turbulent flow farther away.

The fact that unsteady airflow is ubiquitous in natural environments raises the fundamental question of whether insects need to actively respond to the full spectrum of unsteady flows that they experience in the natural world, or whether some of these disturbances can be handled with minimal energetic cost via passive mechanisms. Several recent studies have quantified the flight trajectories and kinematics of insects subjected to various flow perturbations, such as discrete gusts (3), von Kármán vortex streets (4, 5), or steady vortices (6). However, kinematics alone do not allow one to easily distinguish between passive body motions induced by a fluid disturbance and active, voluntary motions generated by the insect to maneuver or correct for a perturbation, particularly when the flow environment is unpredictable or unknown.

In this study, we focus on disentangling the active vs. passive components of a bumblebee's motion while flying through the unsteady, but structured, von Kármán vortex street generated behind a circular cylinder in flow. Bumblebees are ideal model organisms for studying insect flight in complex airflow, since they are relentless foragers that continue to fly in a wide range of weather conditions (7, 8). Prior work has shown that many flying animals, including bumblebees, are most sensitive to aerial disturbances along the lateral axis and rotations along the body's longitudinal axis (i.e. roll). Rolling motion and lateral translation are also critical components of voluntary maneuvering, and some insects have been shown to employ roll as the preferred means to perform rapid flight maneuvers (9, 10). We therefore chose to generate a von Kármán vortex street behind a vertical cylinder, which provides the greatest lateral and rolling perturbations (5), and chose a cylinder size (d = 25 mm) corresponding to a typical bumblebee's wing span, which produces a chain of vortices shed at a rate of 23 Hz (5). We compare the rolling and lateral flight dynamics of bumblebees flying in the vortex street to their flight in steady flow. The relative importance of active control mechanisms vs. passive dynamics could most readily be tested by varying the level of active control that an insect can employ as it flies through unsteady flow, and quantifying the resulting change in its body dynamics and flight trajectory. However, such manipulations are unfeasible in living specimens, but numerical simulations of flying insects offer a suitable alternative. Simulations that combine flapping wing aerodynamics and free-flight dynamics provide a powerful tool that allows for simultaneous manipulation of all kinematic parameters of the insect model, aerodynamic forces and torques, and the ambient flow field. These types of simulations have provided significant insight into both the fine scale flow structures that form over flapping wings and the overall dynamics of insect flight, see (11, 12) for review, but such simulations have only been performed in still air hitherto. Here, we develop a model of a flapping bumblebee flying through a von Kármán vortex street that is nominally identical to the one utilized in experiments on real bees. The model accounts for flow properties at all length scales in the oncoming flow and resolves the resulting lateral and roll flight dynamics of the insect due to its interaction with the wind. Such high fidelity simulations require extensive computational resources, and over 8192 cores and >280 million grid points were used in this study, the highest number that has been used to resolve flapping flight to date.

We compare the flight trajectories and body motions of live bumblebees flying in the wake of a cylinder to those of the model bee in similar flow conditions with no active control, to parsimoniously estimate the role of active control and identify the control strategies being implemented by flying bumblebees.

## Results

### Bumblebee Flight Dynamics

During free flight in both steady and unsteady flows, the flight path of the bees typically consisted of large-amplitude motions at low frequencies (i.e., "casting") that combined lateral translation and rolling (Fig. 1a; (5)). In unsteady flow, higher frequency translational and rolling motions were superimposed on top of these low-frequency casting motions (Fig. 1c, Supplementary Video 1), and earlier investigations revealed that these high-frequency motions occur primarily at the 23-Hz which is the vortex shedding frequency of the cylinder (5). In contrast, the lateral casting movements typically occur at frequencies below 10 Hz, and this was used as the cut-off frequency for examining the low- vs. high-frequency components of motion. The overall results were found to be nominally insensitive to the choice of cut-off frequency, so long as it was smaller than the frequency of the von Kármán street.

The magnitude of lateral accelerations associated with low-frequency, casting motions was similar in both steady and unsteady flow ($p$ = 0.12, Fig 2a). High-frequency motions (> 10 Hz) were not observed in steady flow. However in unsteady flow, the lateral accelerations associated with high-frequency motions were significantly higher than those associated with casting motions ($p < 10^{-7}$, Fig. 2a). The mean absolute roll angles of bees were also similar in both airflow conditions during casting ($p$ = 0.48, Fig. 2a). The absolute roll angles measured at higher frequencies in unsteady flow were significantly smaller than those at lower frequencies ($p < 10^{-6}$, Fig. 2b).

During low-frequency casting, there was a strong, positive correlation between the instantaneous roll angle of bees and their lateral acceleration ($C$ = 0.89±0.16 in steady flow, $C$ = 0.81±0.16 in unsteady flow), and the strength of this correlation was similar in steady and unsteady flow ($p$ = 0.20). However, for higher-frequency motions in unsteady flow, there was a strong, negative correlation between roll angle and lateral acceleration ($C$ = - 0.81±0.13), which was significantly different from the correlation for low-frequency motion in unsteady flow ($p < 10^{-17}$, Fig. 2c).

Due to the high cross correlations, to estimate the relationship between roll angle and lateral acceleration, we performed a linear regression between these variables, for low-frequency casting motions in both flow conditions and for high-frequency motions in unsteady flow (Fig. 1b, 1d, 1e). We found the slope of the regression across all flight trials and compared mean values between flow conditions. For low-frequency casting motions, a statistically similar, positive regression slope was present in both wind conditions ($p$ = 0.866, Fig. 2d). However, at higher frequencies in unsteady flow, the slope of this relationship was negative, and significantly different from the slope for low-frequency motions in unsteady flow ($p < 10^{-14}$, Fig. 2d).

**Assessing the Presence of Active Flight Control**

The high-fidelity numerical simulations were conducted to evaluate the passive flight dynamics of a flapping bumblebee with no active control, flying in a von Kármán vortex street with the same characteristics as the experimental unsteady flow condition. Flight was simulated over 0.5 sec (76 wing beats), during which time the bumblebee interacted with a number of vortices in the von Kármán street (Fig. 3, Supp. Video 2, Supp. Fig. 1). The main outcome of the numerical simulation is the time evolution of the aerodynamic forces and torques acting on the insect, as well as the resulting lateral and rolling motions of the 'numerical' bee. Because the simulations did not include voluntary low-frequency casting motions, the flight dynamics of the model bee in the simulations were high-pass filtered and analyzed in the same way as the high-frequency motions of real bees flying in unsteady flow.

The high-pass filtered lateral and rotational positions and accelerations from the simulation showed good agreement with experimental measurements (Fig. 4 a - d). The time-varying lateral and rolling motions were nominally within the standard deviations of the high-frequency components of motion measured on freely flying bees in unsteady wind, and mean values of absolute roll angle and lateral acceleration were not significantly different from measurements on real bees (roll angle: $p$ = 0.65; lateral acceleration: $p$ = 0.37).

**Discussion**

Aerial locomotion at small scales is very changeling and insects utilize a variety of unsteady aerodynamic mechanisms to generate the necessary forces to not only stay aloft but also perform coordinated maneuvers. In the natural environment volant organisms are also posed with wind unsteadiness. What are the added challenges imposed by wind unsteadiness and what mechanisms do insects employ to mitigate them?

**Low-frequency motions**

In both steady and unsteady winds, bumblebees performed casting movements consisting of large, lateral displacements at low frequencies (Fig. 1a & Supp. Video 1). Such casting motions have also been observed in other insects, including wasps and honeybees (13, 14). While the significance of casting is not completely understood, some insects such as wasps perform casting flights to aid visual processing (13). In the experiments conducted here, the high, positive cross-correlation between roll angle and lateral acceleration at the low frequencies associated with casting behavior (Fig. 2c) suggests a causative relationship between these motions.

The "helicopter model" of flight control, in which the net aerodynamic force vector remains fixed relative to the body and body attitude is adjusted to alter the direction of motion, has been demonstrated in several species of insects and birds (15, 16). Based on this model, acceleration or deceleration along the longitudinal axis requires a change in body pitch, and lateral maneuvers require a body rotation around the roll axis, to reorient the aerodynamic force vector. Our data on low-frequency casting motions in bumblebees supports the helicopter model of flight control. Considering level translation only along the lateral axis, the relationship between body roll ($\psi$, in rad) (refer to Supp. Fig. 1 for coordinate system) and lateral acceleration based on the helicopter model is ($g \tan \psi$) (refer S2.1 for derivation), where g is the acceleration due to gravity. The roll angles observed during casting were sufficiently small, and upon applying the small angle approximation, the relationship between the roll angle and lateral acceleration becomes linear with the proportionality coefficient nominally equal to the acceleration due to gravity (see S2.1 for derivation). In our experiments, since the bees maintained altitude while casting in both steady and unsteady flow, the slope of the regression line between body roll and lateral acceleration was close to the expected value of 9.8 $m/s^2$/rad (steady airflow = 8.6 ± 1.2 $m/s^2$, unsteady airflow = 8.8 ± 1.5 $m/s^2$; Fig. 2d).

The high correlation and positive relationship between low-frequency rolling motions and lateral acceleration in both steady and unsteady winds (Fig. 2d) suggests that the bees were performing the same type of controlled maneuvers at low frequencies in both flow situations. Additionally, the presence of casting motions even in the absence of external flow perturbations (i.e., in steady flow) indicates that these motions were most likely voluntary (see Fig. 5a and Supp. Video 3 for schematic representation and animation of casting motion, respectively). Based on the regression between lateral acceleration and roll, bees are able to generate a lateral acceleration of around 1 $m/s^2$ (typical accelerations measured during casting) by rolling approximately 6 degrees towards the direction of intended motion.

**High-frequency motions**

In the unsteady flow condition, we observed high-frequency rolling motions and lateral accelerations superimposed on top of the slow casting motions, with a clear separation between these two types of oscillations in terms of their temporal scale (Fig. 1c and Supp. Video 1). These high-frequency oscillations in unsteady flow have previously been shown to occur primarily at the vortex-shedding frequency associated with the wake of the cylinder (5). The most obvious difference between the low-frequency casting motions and the higher-frequency oscillations is in the phase relation between rolling and lateral motion. During casting in both flow conditions, we found a strong, positive correlation between roll and lateral acceleration (i.e., the bee rolls to the right and accelerates to the right), whereas at higher frequencies, there is an equally strong but negative correlation between these variables (Fig. 2c). The slope of the regression line between lateral acceleration and roll angle is negative and much larger in magnitude for the high-frequency motions (−80±17 $m/s^2$/rad; Fig. 2d); thus, mean lateral accelerations are significantly larger and mean roll angles significantly smaller for high- vs. low-frequency motions (Fig. 2). These findings suggest that the high-frequency oscillations are fundamentally different from the voluntary, low-frequency casting motions.

The high-frequency oscillations corresponding to the vortex shedding rate could be purely passive motions induced by the external airflow, or they could represent a combination of passive, externally-induced motions and active, corrective maneuvers. To distinguish between these hypotheses, we performed high-fidelity numerical simulations based on the assumption that if the high-frequency motions were purely passive, then a 'numerical' bumblebee with no flight control would exhibit similar motion dynamics as measured in the experiments. The close similarity in temporal patterns of lateral and rolling motions exhibited by real and simulated bumblebees over several cycles of flow disturbance (approximately 20 wing beats; Fig. 4 a-d) lends support to the notion that the high-frequency motions are indeed predominantly due to passive interactions with vortices in the von Kármán street. This suggests that bees do not actively respond to these flow perturbations on a wingbeat-by-wingbeat basis.

To further illustrate the mechanics of the passive, high-frequency motions that bees exhibit in unsteady flow, we propose an idealized "sailboat" model. The governing equations for the high-frequency motions exhibited by the insect can be derived by considering a 'numerical' bumblebee subjected to a periodic lateral wind disturbance. This is a reasonable comparison since the von Kármán street induces similar lateral disturbances on a bee inflight. The relationship between the lateral acceleration and roll angle experienced would be as follows (see S2.2 for derivation):

$$G = \frac{F_y}{m\psi} = \frac{-4\pi^2 f_v^2 I_{xx}}{\bar{L}m} \quad (Eq. 1)$$

Here; $G$ is the relation between lateral acceleration and roll angle ($\psi$, in rad), m is the mass of the bee, $F_y$ is the lateral acceleration, $I_{xx}$ is the roll moment of inertia, $\bar{L}$ is the moment arm, or the distance between the center of area and center of mass of the bee, and $f_v$ is the frequency of the von Kármán street. The model asserts that the force induced by a lateral wind on the wings (which are located above the center of mass for much of the stroke cycle), and the misalignment between the center of pressure and center of mass of the body, leads to a lateral displacement and a corresponding roll, akin to a sailboat rolling when subjected to a side wind (see Fig. 5b and Supplementary Video 4). This model preserves the phase relationship between roll angle and lateral acceleration while buffeting in the experiments. By substituting parameters for the "mean" bumblebee from the experiments into Eq. 1 the proportionality between the lateral acceleration and the roll angle due to the wind disturbance is around -79 m/s$^2$/rad (see S2.2 for elaboration), which is comparable to that measured in the free flying bees (-74 ± 13 m/s$^2$/rad), see Fig. 2d. In Eq. 1 the distance between the center of aerodynamic pressure and center of mass critically influences the relationship, thus the dynamics of the high frequency motion may differ vastly between insects.

**Bimodal Flight Control**

While foraging in the outdoor environment, flying organisms are likely to be subjected to wind induced disturbances over a wide range of frequencies and scales, however the magnitude of fluctuations over different frequencies and scales is still unclear due to lack of outdoor measurements. Nevertheless, performing active corrections for disturbances at all time scales is not only infeasible due to sensorimotor delays but also suboptimal. In this respect, it is vital to optimize the flight control system such that behaviorally relevant tasks can continue to be performed with minimal trade off. In the context of insect flight through unsteady winds, we distinguish between two distinct modes, i.e., the active and passive. Passive interactions with high frequency disturbances may be a suitable strategy since they impart relatively small net displacements that are likely to average over shorter time scales – as in this study where the high frequency displacements were smaller and tended to average over a few centiseconds, see Fig 1c & 5. Do the bees perform any corrective maneuvers when flying in the von Kármán street? The unfiltered time histories of the numerical simulations reveal that though the high frequency motions were relatively small in magnitude, the unsteady wind also introduces instabilities at lower frequencies, see SF2, and if uncorrected, these instabilities tend to increase over time. Therefore to maintain stable flight, some level of flight control is necessary but at frequencies much lower than the frequency of the von Kármán street. The von Kármán street induced disturbances at 23 Hz or 43 ms, which is within the perceptive limits and realms of sensorimotor response times of insects (3). Fruitflies subjected to mechanically induced rapid roll perturbations responded within 5 ms with active changes in wing kinematics (17). The passive response to higher frequencies disturbances noted in our experiments could be attributed to the relatively smaller magnitude of perturbations presented as well as the vast difference in organism properties. Additionally, insect responses to aerial perturbations have been shown to be context-specific whereby escape responses are mediated by specialized neural controllers that can be triggered within much smaller time scales (10).

Combination of active and passive modes in unsteady fluid environments is not uncommon in the animal kingdom. Classical experiments on fish swimming in von Kármán streets reveal a characteristic locomotion pattern knows as Kármán gaiting where fish passively comply with the vortex street resulting in a combined bending and swaying of the body (18). This passive mechanism significantly reduces sensorimotor demands however additional active stabilization is still necessary but at much longer timescales, see (19) for review. The flight trajectories of the bees here do not resemble Kármán gaiting, but the mechanism may be considered analogous. The sensorimotor basis for the implementation of such mechanisms in insects is still an open question while we have a better understanding of aquatic locomotion.

Measurements made on Kármán gaiting fish reveal relatively limited additional cost of locomotion in unsteady fluids (19) but the energetic cost of flight in steady vs unsteady wind is still unclear. However unlike in aquatic locomotion, the variations in orientation angles while maneuvering and high frequency motions inflight would result in a reduction in altitude due to the reorientation of the aerodynamic force vector. Thus apart from implementing corrections to maintain body posture in unsteady wind, an overall increase in force production may be necessary to maintain altitude. Additionally performing active corrections may impose further metabolic cost but measurements on bumblebees in rarified medium and in load-lifting tests reveal that they can increase force production with relatively limited energetic expense (20, 21). Taken together we observe a novel flight control mechanism implemented by bumblebees to cope with complex and unpredictable winds, a mechanism that could be widespread among volant organisms and one that could be suitable for artificial miniature aerial vehicles flying in outdoor environments.

## MATERIALS AND METHODS

### Study specimens

Bumblebees (*Bombus impatiens*) from a commercial breeder (BioBest) were maintained in the lab and given continuous access to a foraging chamber, where they could feed freely from an artificial flower containing linalool-scented nectar. Nominally similar sized bumblebees (body length = 14 mm ± 0.5 mm, mass = 165 mg ± 10%) were selected for flight experiments. The bees were placed in a transparent chamber (0.4 x 0.4 x 0.4 m) without access to food, for approximately two hours prior to the experiment

### Flight Tests

All experiments were conducted in a 6 m-long, suction-type, open-return wind tunnel with a 0.9 x 0.5 x 0.5 m working section. Once sufficiently starved, each bee was placed in the wind tunnel (with no airflow) where it could feed from an artificial flower resembling the one in the foraging chamber. Once feeding commenced, the bee was allowed to feed for approximately 10 seconds, then separated from the nectar source and released at the downstream end of the wind tunnel. Upon release, if the bee did not fly towards the artificial flower, it was manually re-introduced to the nectar source and subsequently separated. This procedure was repeated until direct flights to the nectar source were observed. Once consistent behavior was established, wind was introduced and bees were filmed as they flew upstream. The wind-speed was set to ~2.55 m/s, which represents an intermediate cruising velocity for bumblebees (8, 22).

Each bee was flown in two airflow conditions, steady and unsteady. In the wind tunnel with unimpeded (steady) flow, a uniform velocity profile was present across the interrogation volume (< 2 % variation in mean flow speed) and turbulence intensity (standard deviation/mean wind speed) was less than 1.2 %. With a cylinder (d = 25 mm) positioned vertically at the inlet of the test section, a von Kármán street developed in the wake and vortex shedding occurred at 23 Hz, in agreement with the predicted vortex shedding Strouhal number of 0.19 (23, 24). The vertical cylinder induced lateral velocity fluctuations at the shedding rate over a range of ± 1.2 m/s. Refer to (5) for further elaboration of flow conditions.

We filmed bees and quantified airflow within a specific interrogation volume (a cube with side lengths of 100 mm, located 100 mm downstream from the cylinder). The downstream distance was chosen to avoid the recirculating region in the near wake of the cylinder and to allow the formation of a periodic von Kármán vortex street. A total of 13 bees were subjected to this assay, and paired trials in steady vs. unsteady flow were obtained for each individual.

*Kinematic reconstruction and analysis*

Prior to experiments, a triangular marker was affixed to the dorsal surface of the thorax of each bee, to aid in quantifying its position and orientation during flight. The markers consisted of three black points representing the vertices of an isosceles triangle (measuring 2.7 x 2.3 mm) set upon a white background. During flight trials, bees were filmed as they flew through the interrogation volume using two Photron SA3 high-speed cameras sampling at 1000 Hz, placed above the wind tunnel at approximately 30° from the vertical. The recorded flight sequences were digitized using an open-source MATLAB-based routine, DLTdv5 (25), utilizing the automated tracking feature to localize the three black points on the triangular marker throughout each sequence. Digitization error in localizing the centroids of marker points is expected to be of the order of 1-2 pixels, which is much smaller than the mean number of pixels separating the markers (~30). The digitized position data were initially passed through a $4^{th}$-order, Butterworth, low-pass filter to remove any higher frequency errors due to the digitization process, with a cutoff frequency of 30 Hz, which is lower than the Nyquist frequency (500 Hz). For analysis of displacements at higher frequencies, the displacements were filtered using a high pass filter with 10Hz cut-off.

Instantaneous velocity and accelerations of the bees were calculated through numeric differentiation of the digitized position, see SF1 for coordinate system. The influence of flow conditions on the body orientation and rotation rates of bees was assessed by evaluating variation in roll, pitch and yaw angles of the triangular markers, using a rigid body assumption. The method detailed in (5) was used to calculate the instantaneous orientation and rotation rates of the bees, and to evaluate the errors associated with data capture and subsequent analysis. Subsequently, during analysis the lateral and rotational velocities and accelerations were separated into low (<10 Hz) and high frequency components of motion by applying a $4^{th}$ order Butterworth 10-Hz low- or high-pass filter.

*Numerical Modeling*

The numerical method employed in this study is described in greater detail in (26, 27), and the bumblebee model has been introduced in (28). In this section we provide a short overview of the model, we refer to Supplementary Figure 1 for further elaboration. The bumblebee is approximated by three rigid elements: the body and two wings, which move with respect to each other. The characteristic size of the insect is the wing length, which is equal to R = 12.5 mm. The wings follow a prescribed periodic flapping motion with frequency f = 152 Hz. Their positional angle varies as $\Phi = 24°+115°/\sin(2\pi ft)$, the elevation with respect to the stroke plane is constant and equal to $\theta=12.55°$. The feathering angle $\alpha$ is equal to 70° during upstroke and -40° during downstroke, with sinusoidal variation at the reversals. The angle between the body longitudinal axis and the horizontal plane is equal to 24.5°, and the angle between the wing stroke plane and the horizontal plane is equal to 28°.

These wing kinematics ensure trimmed flight at $u_\infty$=2.5 m/s, assuming that the body mass is 175 mg. The position of the insect varies in time according to the solid-body dynamics model (26). In this study, only two degrees of freedom are taken into account:

lateral displacement and roll rotation about the longitudinal axis of the body. The roll moment of inertia is equal to $I_{xx}$=10.92 x $10^{-10}$ kg m². As with the experimental data gathered from bees in unsteady flow, the calculated translational and rotational positions and accelerations of the simulated bee were initially filtered using 4th order Butterworth low-pass filter with cut-off frequency 30 Hz. Subsequently to isolate the higher frequency motions, the data was high-pass filtered using a 4th order Butterworth 10 Hz high-pass filter.

To generate unsteady flow in the simulation, a vertical cylinder (diameter=26.4 mm, Re = 4200) is placed in front of the insect, at approximately 80 mm from the insect's center of mass (Fig. 3, Supplementary Fig. 1). The cylinder has a salient detail on one side in order to break the symmetry of the flow. The flow is governed by the incompressible Navier-Stokes equations, and the surrounding air has density $\rho$=1.177kg/m³ and kinematic viscosity $\eta$ = 1.57x$10^{-5}$ m²/s. The flow domain is a rectangular channel of length 132 mm, having a 105.6 x 52.8 mm cross-section. The no-slip boundary condition at the surfaces of the insect and cylinder as well as the flow outlet condition, are imposed using the volume penalization method (see (26, 27) for details). The penalized equations are solved numerically using a Fourier pseudo-spectral method. The boundary condition on the side-walls of the channel is periodic. The flow domain is discretized using a uniform Cartesian grid consisting of 960 x 768 x 384 points (> 280 million grid points). The volume penalization parameter is equal to $\eta$=$10^{-3}$. The resulting simulated flow environment accurately reproduced the unsteady flow environment used in experiments, with a vortex shedding rate and transient velocities matching those measured in the experiments.

*Statistical Analysis*

Statistical significance of experimental results was analyzed by performing paired t-tests (n = 14 individuals in all cases) in MATLAB, between the low-frequency components of flight in steady vs. unsteady flow, and between the low- and high-frequency components of flight in unsteady flow. The mean lateral and rotational displacement and accelerations obtained from the simulation were compared to those measured on freely flying bees in the experiments by performing a one-sample t-test.

**References**

1.  Stull RB (1988) *An Introduction to Boundary Layer Meteorology* ed Stull RB (Springer Netherlands, Dordrecht) doi:10.1007/978-94-009-3027-8.

2.  Watkins S, Milbank J, Loxton BJ, Melbourne WH (2006) Atmospheric Winds and Their Implications for Microair Vehicles. *AIAA J* 44(11):2591–2600.

3.  Vance JT, Faruque I, Humbert JS (2013) Kinematic strategies for mitigating gust perturbations in insects. *Bioinspir Biomim* 8(1):016004.

4.  Ortega-Jimenez VM, Greeter JSM, Mittal R, Hedrick TL (2013) Hawkmoth flight stability in turbulent vortex streets. *J Exp Biol* 216(Pt 24):4567–79.

5.  Ravi S, Crall JD, Fisher A, Combes SA (2013) Rolling with the flow: bumblebees flying in unsteady wakes. *J Exp Biol* 216(Pt 22):4299–309.

6. Ortega-Jimenez VM, Mittal R, Hedrick TL (2014) Hawkmoth flight performance in tornado-like whirlwind vortices. *Bioinspir Biomim* 9(2):025003.

7. Crall J, Combes S (2013) Blown in the wind: Bumblebee temporal foraging patterns in naturally varying wind conditions. *Integrative and Comparative Biology* (OXFORD UNIV. PRESS INC), p E270.

8. Riley JR, et al. (1999) Compensation for wind drift by bumble-bees. *Nature* 400(6740):126–126.

9. Dudley R (2002) Mechanisms and implications of animal flight maneuverability. *Integr Comp Biol* 42(1):135–40.

10. Muijres FT, Elzinga MJ, Melis JM, Dickinson MH (2014) Flies evade looming targets by executing rapid visually directed banked turns. *Science* 344(6180):172–7.

11. Shyy W, Aono H, Kang C-K, Liu H (2013) *An Introduction to Flapping Wing Aerodynamics* (Cambridge University Press). Volume 37.

12. Sun M (2014) Insect flight dynamics: Stability and control. *Rev Mod Phys* 86(2):615–646.

13. Boeddeker N, Dittmar L, Stürzl W, Egelhaaf M (2010) The fine structure of honeybee head and body yaw movements in a homing task. *Proc Biol Sci* 277(1689):1899–906.

14. Braun E, Dittmar L, Boeddeker N, Egelhaaf M (2012) Prototypical components of honeybee homing flight behavior depend on the visual appearance of objects surrounding the goal. *Front Behav Neurosci* 6:1.

15. Thomas ALR, Taylor GK (2001) Animal Flight Dynamics I. Stability in Gliding Flight. *J Theor Biol* 212:399–424.

16. Ros IG, Bassman LC, Badger MA, Pierson AN, Biewener AA (2011) Pigeons steer like helicopters and generate down- and upstroke lift during low speed turns. *Proc Natl Acad Sci U S A* 108(50):19990–5.

17. Beatus T, Guckenheimer JM, Cohen I (2015) Controlling roll perturbations in fruit flies. *J R Soc Interface* 12(105):20150075–20150075.

18. Liao JC, Beal DN, Lauder G V, Triantafyllou MS (2003) Fish exploiting vortices decrease muscle activity. *Science* 302(5650):1566–9.

19. Liao JC (2007) A review of fish swimming mechanics and behaviour in altered flows. *Philos Trans R Soc Lond B Biol Sci* 362(1487):1973–93.

20. Dillon ME, Dudley R (2014) Surpassing Mt. Everest: extreme flight performance of alpine bumble-bees. *Biol Lett* 10(2):20130922.

21. Hedenstrom A, Ellington CP, Wolf TJ (2001) Wing wear, aerodynamics and flight energetics in bumblebees (Bombus terrestris): an experimental study. *Funct Ecol*

15(4):417–422.

22. Ellington CP (1991) Limitations on animal flight performance. *J Exp Biol* 160:71–91.

23. Roshko A (1961) Experiments on the flow past a circular cylinder at very high Reynolds number. *J Fluid Mech* 10(03):345.

24. Vickery BJ (1966) Fluctuating lift and drag on a long cylinder of square cross-section in a smooth and in a turbulent stream. *J Fluid Mech* 25(03):481.

25. Hedrick TL (2008) Software techniques for two- and three-dimensional kinematic measurements of biological and biomimetic systems. *Bioinspir Biomim* 3(3):034001.

26. Engels T, Kolomenskiy D, Schneider K, Sesterhenn J (2015) FluSI: A novel parallel simulation tool for flapping insect flight using a Fourier method with volume penalization. *arXiv* 1506.06513.

27. Kolomenskiy D, Schneider K (2009) A Fourier spectral method for the Navier–Stokes equations with volume penalization for moving solid obstacles. *J Comput Phys* 228(16):5687–5709.

28. Engels T, Kolomenskiy D, Schneider K, Lehmann F-O, Sesterhenn J (2016) Bumblebee Flight in Heavy Turbulence. *Phys Rev Lett* 116(2):028103.

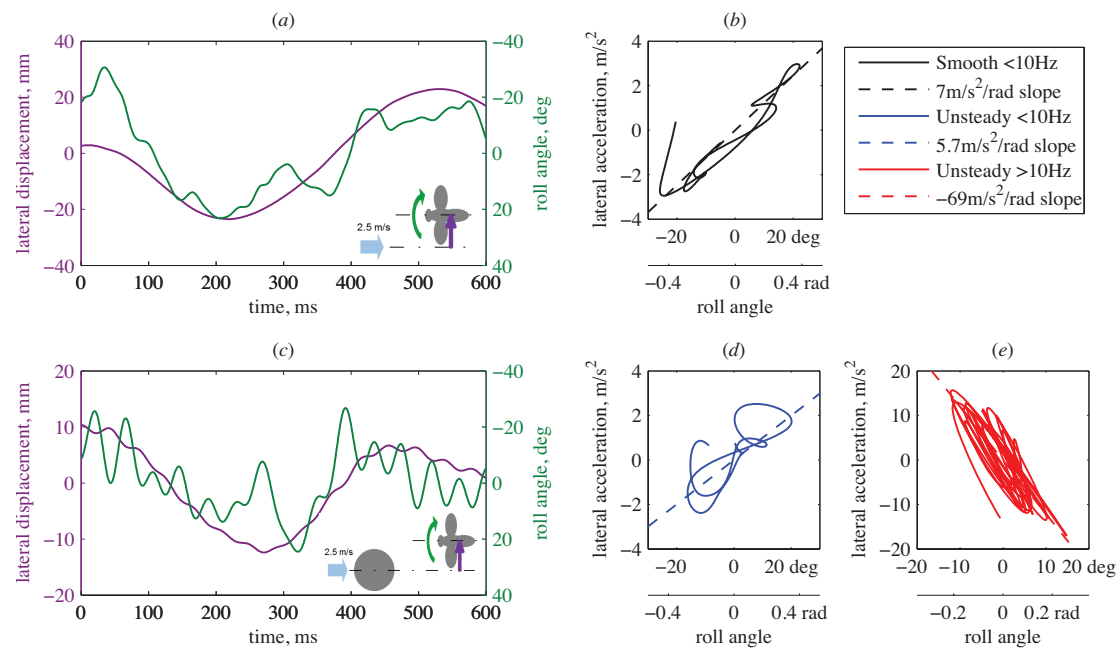

*Figure 1: (a) Sample time series of lateral displacement and roll angle of a freely flying bumblebee in steady flow. (b) Instantaneous lateral acceleration vs. roll angle for lower-frequency (<10Hz) motions of the flight shown in (a). (c) Sample time series of lateral displacement and roll angle for the same bee flying in the*

*unsteady wake of a vertical cylinder. (d) Instantaneous lateral acceleration vs. roll angle for lower- (d) and higher-frequency (e) motions of the flight shown in (c). In (b), (d) and (e) the linear regression between lateral acceleration and roll angle is shown by a dashed line.*

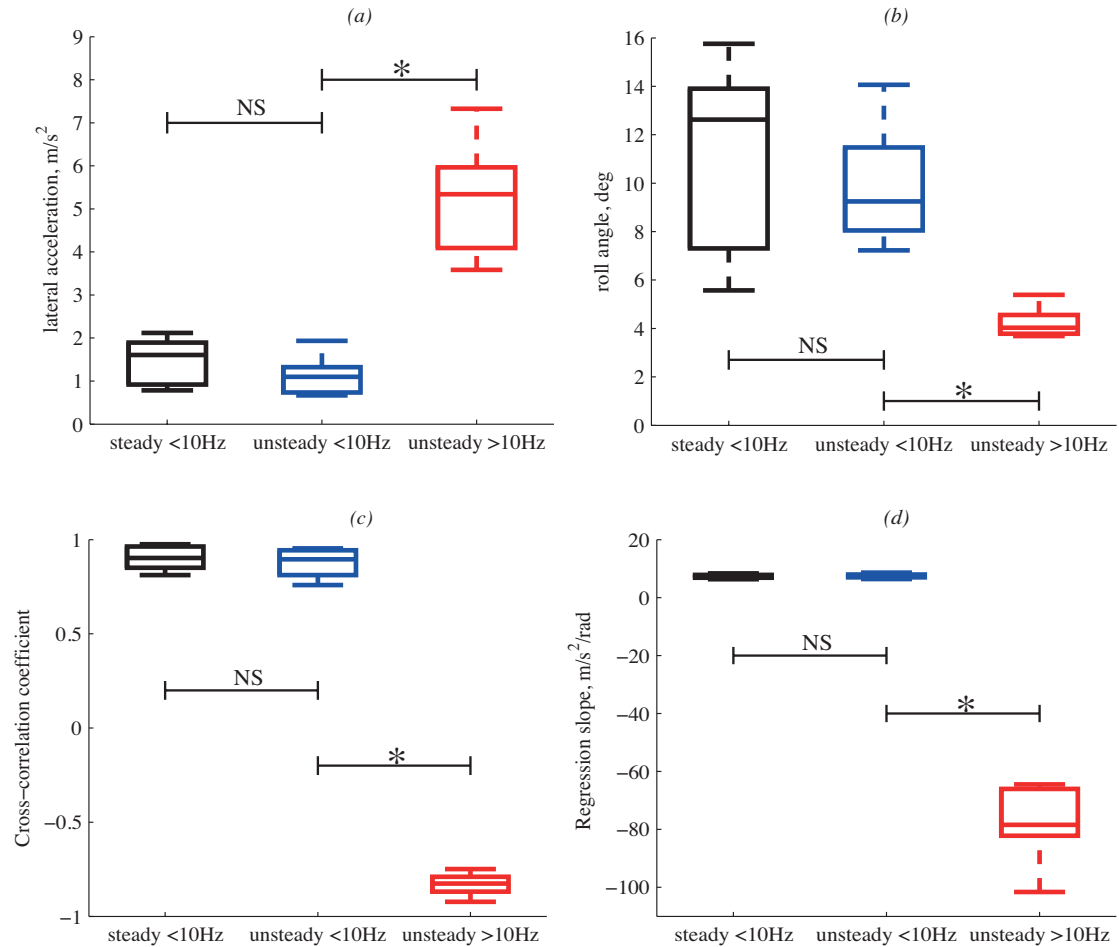

*Figure 2: Mean absolute lateral acceleration (a) and (b) roll angle of lower-frequency (<10 Hz) casting motions in steady and unsteady flow, and higher-frequency (>10 Hz) motions in unsteady flow. (c) Correlation coefficients (zero time-lag, normalized cross correlation) for the relationship between instantaneous lateral acceleration and roll angle for lower-frequency (<10 Hz) casting motions in steady and unsteady flow, and higher-frequency (>10 Hz) motions in unsteady flow. (d) The slope of the regression line between lateral acceleration and roll angle, in the same flow conditions as in (c). Asterisks indicate a significant difference at the $p < 0.05$ level, n = 14 for all conditions.*

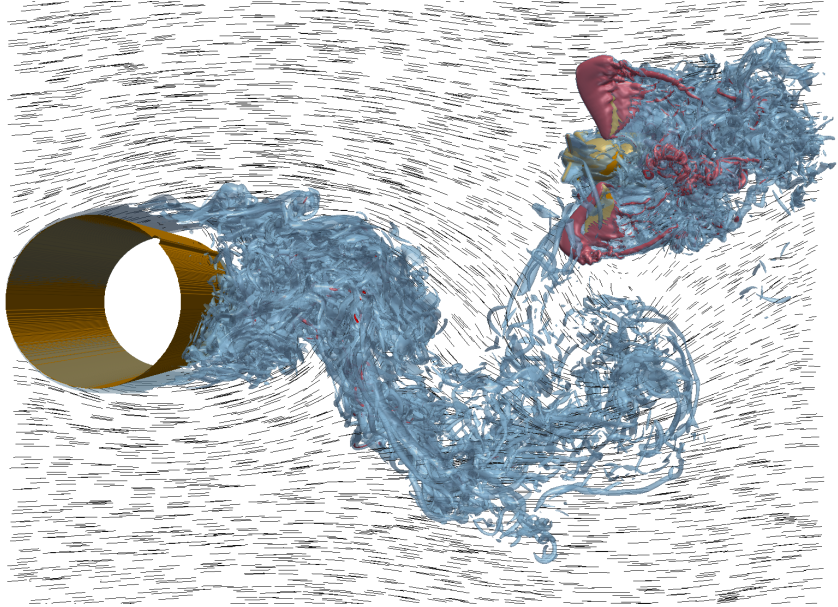

*Figure 3: Birds eye view of the flow field obtained with numerical simulation. Isosurfaces of dimensionless vorticity magnitude are shown as, light blue |ω| = 5, red |ω| = 40. Arrows show the velocity in the horizontal plane that passes through the insect's center of gravity. The insect is colored in orange. Flow is shown at t = 106.5 ms after the beginning of the simulation.*

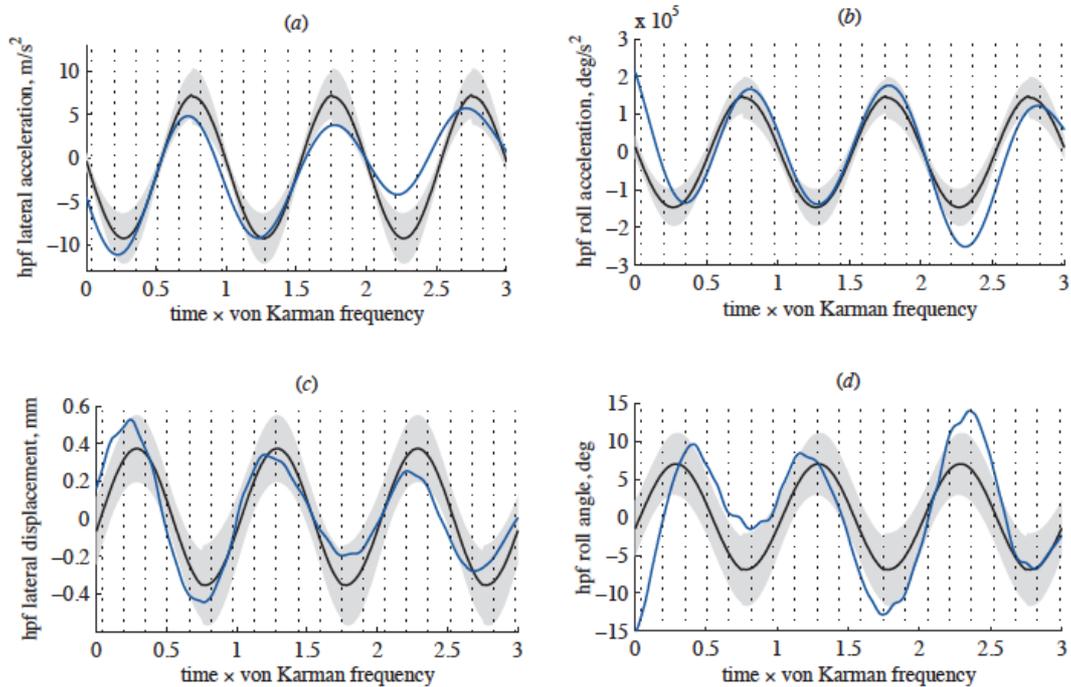

*Figure 4: Time evolution of high pass filtered (hpf, > 10Hz) lateral accelerations (a) and displacements (b), and of roll accelerations (c) and angles (d) from the numerical simulation and experiments on freely flying bees. Blue line shows results of the numerical simulation (for motions > 10 Hz), and black line/gray shading*

*show the mean and standard deviation of measurements from experiments in unsteady flow (motions > 10 Hz). Dashed vertical lines represent wing beat periods.*

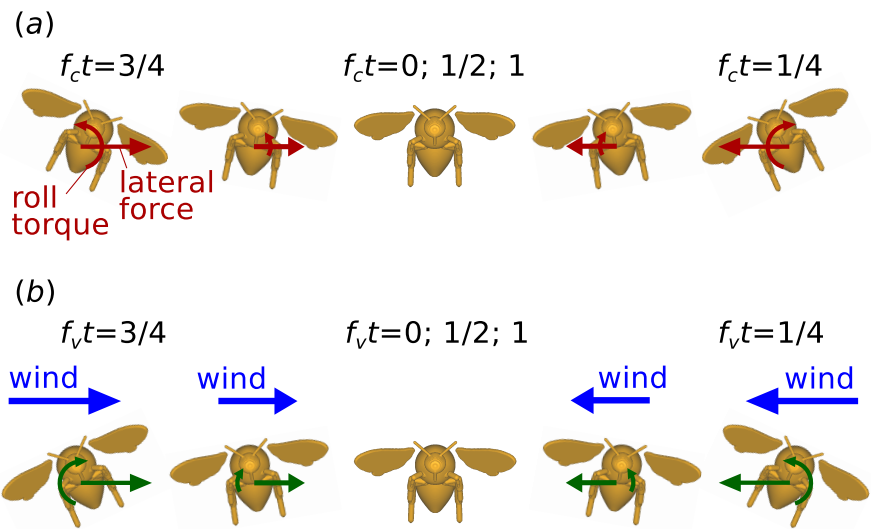

*Figure 5: Schematic representation of the relationship between lateral and roll, accelerations and displacements, during (a) low frequency casting maneuvers (helicopter model) and (b) while high frequency buffeting in unsteady winds (sailboat model). The red and green arrows in (a) and (b) respectively represent the self-initiated and wind-induced forces and torques respectively. Here $f_c$ (< 10 Hz) and $f_v$ (≈ 23 Hz) are casting and von Kármán frequencies respectively and t is normalized time.*